\documentclass[nofootinbib,twocolumn,showpacs,preprintnumbers,pre,aps]{revtex4-1}

\usepackage{graphicx}
\usepackage{bm}
\usepackage{amsmath}
\usepackage{amssymb}
\usepackage{color}
\usepackage{natbib}
\usepackage{mathrsfs}

\begin{document}

\title{
Non-reciprocal response of a two-dimensional fluid with odd viscosity
}

\author{Yuto Hosaka}

\author{Shigeyuki Komura}\email{komura@tmu.ac.jp}

\affiliation{
Department of Chemistry, Graduate School of Science,
Tokyo Metropolitan University, Tokyo 192-0397, Japan
}

\author{David Andelman}\email{andelman@tauex.tau.ac.il}

\affiliation{
Raymond and Beverly Sackler School of Physics and Astronomy, Tel Aviv University, Ramat Aviv, Tel Aviv 69978, Israel}


\begin{abstract}
We discuss the linear hydrodynamic response of a two-dimensional active chiral compressible fluid with odd viscosity.
The viscosity coefficient represents broken time-reversal and parity symmetries in the 2D fluid and characterizes the deviation of the system from a passive fluid.
Taking into account the hydrodynamic coupling to the underlying bulk fluid, we obtain the odd viscosity-dependent mobility tensor, which is responsible for the non-reciprocal hydrodynamic response to a point force.
Furthermore, we consider a finite-size disk moving laterally in the 2D fluid and demonstrate that the disk experiences a non-dissipative lift force in addition to the dissipative drag one.
\end{abstract}

\maketitle

\section{Introduction}
\label{sec:intro}

Two-dimensional (2D) active chiral fluids have been predicted to have a new rheological property called \textit{odd viscosity}~\cite{avron1998}.
Over two decades ago, Avron has shown that when time-reversal and parity symmetries are broken, the viscosity tensor of a 2D isotropic fluid could have an anti-symmetric (odd) part that does not result in dissipation~\cite{avron1998,banerjee2017}.
The origin of the odd viscosity is explained by coarse-grained theories, such as Onsager's reciprocal relations~\cite{avron1998} and Green-Kubo relations for viscosity coefficients~\cite{epstein2020,hargus2020,han2020}.
Furthermore, in microscopic approaches, it was shown that active chiral fluids composed of self-spinning objects also exhibit odd viscosity~\cite{banerjee2017,markovich2020}.

In order to observe odd viscosity in physical systems, several protocols have been proposed.
For incompressible fluids, it was predicted that odd viscosity can emerge at a dynamical boundary that is subjected to no-stress boundary conditions~\cite{ganeshan2017,abanov2018}.
Experimentally, Soni \textit{et al.}\ measured odd viscosity in active chiral fluids by observing the boundary dynamics of a fluid~\cite{soni2019}. 
Odd viscosity has also been measured by using molecular dynamics simulations and its Green-Kubo representation~\cite{hargus2020,han2020}.

In addition, more fundamental phenomena in active chiral systems, such as responses to point forces or finite-size bodies, have been studied theoretically~\cite{ganeshan2017,khain2020,braverman2020,scheibner2020,reichhardt2019,kogan2016}.
Recently, the non-reciprocity of the point force response was investigated in a 3D anisotropic fluid with odd viscosity~\cite{khain2020} as well as in an active solid material~\cite{braverman2020,scheibner2020}.
As for the finite-size body response, lift force was observed in active chiral granular media~\cite{reichhardt2019} and in a 2D fluid with inertia~\cite{kogan2016}, whereas no such force was found in the incompressible limit~\cite{ganeshan2017}.
Despite these intensive findings, very little is known about the linear hydrodynamic response of a 2D active chiral fluid.

In this paper, we discuss the hydrodynamic response of a 2D isotropic compressible fluid with odd viscosity, which can be regarded as a 2D active chiral fluid~\cite{avron1998,banerjee2017,epstein2020}.
Taking into account the 3D bulk fluid coupled to the 2D fluid layer and employing the lubrication approximation for the 3D fluid~\cite{barentin1999,barentin2000,elfring2016}, we analytically obtain the asymmetric mobility tensor of the 2D fluid in the presence of odd viscosity.
Because of such a non-reciprocal hydrodynamic response,  a perpendicular fluid flow develops and breaks the axial symmetry of the flow with respect to the driving force.
Extending the point force response, we derive viscous forces on a rigid disk that moves laterally in the 2D active chiral fluid.
As a consequence of the non-reciprocal hydrodynamic response due to odd viscosity, we find that lift force acts on the driven disk.

There are two reasons that motivated us to consider a 2D compressible fluid with odd viscosity in contact with an underlying 3D fluid. 
As pointed out in Ref.~\cite{ganeshan2017}, the velocity field is independent of the odd viscosity in a 2D incompressible fluid, when non-slip boundary conditions are imposed on a moving object. 
This means that the effects of odd viscosity cannot be directly seen in a 2D incompressible fluid. 
Moreover, for a pure 2D fluid at low Reynolds number, a linear relation between the velocity and the viscous drag force acting on a translating disk cannot be obtained~\cite{lamb1975,Landau1987}.
This is known as the Stokes paradox that originates from the constraint of momentum conservation in a pure 2D system.
This paradox can be resolved, e.g., by considering momentum decay to an underlying 3D fluid~\cite{evans1988,diamant2009,oppenheimer2009,oppenheimer2010,ramachandran2010,ramachandran2011}.

In the next section, we review the concept of the odd viscosity~\cite{avron1998,banerjee2017,epstein2020}.
Then, we introduce in Sec.~\ref{sec:model} the hydrodynamic equations for the 2D active chiral fluid layer by taking into account the coupling to the underlying 3D bulk fluid~\cite{barentin1999,barentin2000,elfring2016}.
In Sec.~\ref{sec:mobility}, we derive the corresponding mobility tensor and calculate the velocity fields induced by a point force or a force dipole.
In Sec.~\ref{sec:force}, we discuss the force and torque acting on a moving disk of a finite size.
Finally,  a summary of our work and some further comments are given in Sec.~\ref{sec:discussion}.

\begin{table*}
\caption{\label{tab:table1}
The components associated with the three viscosity coefficients $\eta_{\rm d}$ (dilatation), $\eta_{\rm s}$ (shear), and $\eta_{\rm o}$ (odd) in the $\eta_{ijk\ell}$ tensor under index permutations [see Eq.~(\ref{eq:eta})].
The $+$ ($-$) sign denotes that the components are symmetric (anti-symmetric) under a given index permutation.
Components that include even (odd) number of $\epsilon_{ij}$ are symmetric (anti-symmetric) under the parity transformation in a 2D system
($x\to-x$, $y\to y$)~\cite{epstein2020}.
}
\begin{ruledtabular}
\begin{tabular}{cccccc}
Viscosity coefficients &Components&$i\leftrightarrow j$&$k\leftrightarrow\ell$&
$ij\leftrightarrow k\ell$&Parity
\\ \hline
$\eta_{\rm d}$&$\delta_{ij}\delta_{k\ell}$&+&+&+&+\\
$\eta_{\rm s}$&$\delta_{i k} \delta_{j \ell}+\delta_{i \ell} \delta_{j k}-\delta_{i j}\delta_{k\ell}$&+&+&+&+\\
$\eta_{\rm o}$&$\epsilon_{i k} \delta_{j \ell}+\epsilon_{j \ell} \delta_{i k}
+\epsilon_{i \ell} \delta_{j k}+\epsilon_{j k} \delta_{i \ell}$&+&+&$-$&$-$
\end{tabular}
\end{ruledtabular}
\label{limits}
\end{table*}

\section{Odd viscosity}
\label{sec:odd}

Here we briefly review the concept of odd viscosity and its contribution to the fluid stress tensor in
two dimensions~\cite{avron1998,banerjee2017,epstein2020}.
First, the strain rate tensor is defined as $v_{k\ell} = (\partial_k v_\ell + \partial_\ell v_k)/2$, where
$v_i$ is the 2D velocity component, $\partial_i =\partial/\partial r_i$ is the 2D differential operator component, and $\mathbf{r}=(x,y)$.
The general linear relation between $v_{k\ell}$ and the fluid stress tensor $\sigma_{ij}$ is given by
\begin{align}
\sigma_{ij} = \eta_{ijk\ell}v_{k\ell},
\label{eq:sigma}
\end{align}
where $\eta_{ijk\ell}$ is the fourth-rank viscosity tensor.
Throughout our work, we assume summation over repeated indices. 
Notice that the choice of the Cartesian coordinates is just done for convenience and is not a requirement.

In an isotropic fluid, rotational invariance of the system requires the symmetry of the stress tensor under the exchange of indices $i\leftrightarrow j$, i.e., $\sigma_{ij}=\sigma_{ji}$.
This enforces the symmetry relation of the viscosity tensor $\eta_{ijk\ell}=\eta_{jik\ell}$, as inferred from Eq.~(\ref{eq:sigma}).
Since $v_{k\ell}$ is a symmetric tensor by definition, the symmetry under the exchange $k\leftrightarrow\ell$ always holds, leading to the symmetry relation $\eta_{ijk\ell}=\eta_{ij\ell k}$.

Extending the above symmetry argument, Avron introduced a new type of index exchange $ij\leftrightarrow k\ell$, which implies time-reversal transformation~\cite{avron1998,banerjee2017,epstein2020}.
In light of such a pair exchange, the viscosity tensor can generally be split into symmetric (even) and anti-symmetric (odd) parts $\eta_{ijk\ell}=\eta_{ijk\ell}^{\rm S}+\eta_{ijk\ell}^{\rm A}$, where $\eta_{ijk\ell}^{\rm S}=\eta_{k\ell ij}^{\rm S}$ and $\eta_{ijk\ell}^{\rm A}=-\eta_{k\ell ij}^{\rm A}$.
The anti-symmetric term $\eta_{ijk\ell}^{\rm A}$ exists as a consequence of broken time-reversal symmetry in a 2D fluid.
Under the assumption that $\sigma_{ij}$ is isotropic, the general viscosity tensor can be written as~\cite{banerjee2017,epstein2020}
\begin{align}
\eta_{i j k\ell} &=
\eta_{\rm d}\delta_{i j}\delta_{k\ell}
+\eta_{\rm s}\left(\delta_{i k} \delta_{j \ell}+\delta_{i \ell} \delta_{j k}-\delta_{i j}\delta_{k\ell}\right)\nonumber\\
&+\frac{1}{2}\eta_{\rm o}
\left(\epsilon_{i k} \delta_{j \ell}+\epsilon_{j \ell} \delta_{i k}
+\epsilon_{i \ell} \delta_{j k}+\epsilon_{j k} \delta_{i \ell}
\right),
\label{eq:eta}
\end{align}
where $\eta_{\rm d}, \eta_{\rm s}$, and $\eta_{\rm o}$ are 2D dilatational, shear, and odd viscosities, respectively,
$\delta_{ij}$ is the Kronecker delta, and $\epsilon_{ij}$ is the 2D Levi-Civita tensor with $\epsilon_{xx}=\epsilon_{yy}=0$ and $\epsilon_{xy}=-\epsilon_{yx}=1$.

The above viscosity tensor $\eta_{ijk\ell}$ is symmetric under the parity transformation in a 2D system ($x\to-x$, $y\to y$)~\cite{epstein2020}, and hence it is parity-even.
This is because both $\sigma_{ij}$ and $v_{k\ell}$ are parity-even in Eq.~(\ref{eq:sigma}).
On the other hand, terms that include odd number of $\epsilon_{ij}$ are parity-odd.
Hence, one concludes from Eq.~(\ref{eq:eta}) that $\eta_{\rm o}$ exists only if both time-reversal and parity symmetries are broken~\cite{banerjee2017}.
In Table~\ref{tab:table1}, the above permutations of the viscosity-tensor components are summarized.

Substituting Eq.~(\ref{eq:eta}) into Eq.~(\ref{eq:sigma}), we obtain the stress tensor of a 2D compressible fluid with odd viscosity as
\begin{align}
\sigma_{ij} &= (\eta_{\rm d}-\eta_{\rm s})\delta_{i j}\partial_kv_k
+\eta_{\rm s}(\partial_j v_i+\partial_i v_j)\nonumber\\
&+\frac{1}{2}\eta_{\rm o}
\left(
\partial_j v_i^\ast+\partial_i v_j^\ast+\partial_j^\ast v_i+\partial_i^\ast v_j
\right),
\label{eq:stress}
\end{align}
where $v_i^\ast \equiv \epsilon_{ik}v_k$ is the velocity vector rotated clockwise by $\pi/2$ and $\partial_i^\ast \equiv \epsilon_{ik}\partial_k$.
Since the odd viscosity does not contribute to the energy dissipation, $(\partial_iv_j)\sigma_{ij}$, the sign of $\eta_{\rm o}$ can be either positive or negative.

\section{Active chiral fluid}
\label{sec:model}

\begin{figure}[tbh]
\begin{center}
\includegraphics[scale=0.65]{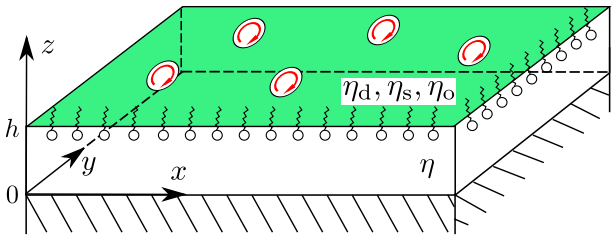}
\end{center}
\caption{
Schematic sketch of an active chiral fluid characterized by odd viscosity.
An infinitely large, flat, and thin 2D fluid layer (green) (e.g., a monolayer formed by amphiphiles) is located at $z=h$ having 2D dilatational, shear, and odd viscosities, $\eta_{\rm d}$, $\eta_{\rm s}$, and $\eta_{\rm o}$, respectively.
The fluid interface is in contact with air $(z>h)$ and a 3D fluid underneath $(0<z<h)$ having a 3D shear viscosity $\eta$.
The 3D fluid is bounded by an impermeable flat wall located at $z=0$, and its velocity is assumed to vanish at $z=0$.
On the 2D fluid layer at $z=h$, both time-reversal and parity symmetries are broken (e.g., due to the self-spinning objects injecting energy into the fluid), giving rise to the possibility of an odd viscosity, $\eta_{\rm o}$, in the 2D layer.
\label{fig:system}
}
\end{figure}

We consider a 2D layer of an active chiral compressible fluid particularly having odd viscosity, which is flat, thin, infinitely large and overlays a 3D bulk fluid (e.g.\ water).
One of the realizations of such a system is schematically depicted in Fig.~\ref{fig:system}.
The bulk fluid has a 3D shear viscosity $\eta$ and is in contact with an impermeable flat wall located at $z=0$, where the fluid 
velocity vanishes.
In order to clearly see the odd viscosity effect~\cite{ganeshan2017}, we suppose that the 2D fluid layer is compressible, so that it has both 2D dilatational and shear viscosities, $\eta_{\rm d}$ and $\eta_{\rm s}$, respectively.
In physical systems, such a fluid can be realized by a monolayer of amphiphiles that are loosely packed on the interface at $z=h$~\cite{barentin1999,barentin2000,elfring2016}.
More details on the physical realization of our model will be discussed in Sec.~\ref{sec:discussion}.

In addition to the above viscosity coefficients, we assume that the 2D layer has odd viscosity, $\eta_{\rm o}$, that is an important measure of how far the fluid departs from passive fluids.
We will not specifically focus on the origin of the odd viscosity, but it can be attributed, for example, to self-spinning objects immersed in the 2D fluid layer that break both time-reversal and parity symmetries~\cite{banerjee2017,souslov2019}.
In this case, one has to assume that the active rotors are homogeneously distributed in the 2D fluid layer and their concentration is small enough.
Under this condition, the 2D layer can be regarded as a layer of a continuum active chiral fluid with a constant 
odd viscosity, $\eta_{\rm o}$.

For the 2D fluid layer introduced above, the momentum balance equation at low Reynolds number can be written as
\begin{align}
-\nabla \Pi+\nabla\cdot\boldsymbol{\sigma}+\mathbf{f}_{\rm b} + \mathbf{F} =0,
\label{eq:balance}
\end{align}
where $\nabla=(\partial_x, \partial_y)$ stands for the 2D gradient operator, $\Pi$ is the 2D hydrostatic pressure, $\boldsymbol{\sigma}$ is the stress tensor given in Eq.~(\ref{eq:stress}) that includes the odd viscosity $\eta_{\rm o}$,
$\mathbf{f}_{\rm b}$ is the force exerted on the 2D fluid layer by the underlying 3D bulk fluid,
and $\mathbf{F}$ is any other force density acting on the 2D fluid.

The bulk underneath the 2D fluid layer is a pure 3D fluid.
We denote its velocity field by the vector $\mathbf{u}(\mathbf{r},z)$ and the 3D hydrostatic pressure by $p(\mathbf{r},z)$.
The corresponding Stokes equation is
\begin{align}
\eta\tilde{\nabla}^2\mathbf{u}-\tilde{\nabla}p &=0,
\label{eq:bulkstokes}
\end{align}
with $\eta$ being the 3D shear viscosity of the bulk fluid and $\tilde{\nabla}$ being the 3D gradient operator.
The incompressibility condition for $\mathbf{u}$ reads
\begin{align}
\tilde{\nabla}\cdot\mathbf{u} &=0.
\label{eq:bulkincomp}
\end{align}

In order to obtain $\mathbf{f}_{\rm b}$, we solve the hydrodynamic equations for the bulk fluid
in Eqs.~(\ref{eq:bulkstokes}) and (\ref{eq:bulkincomp})~\cite{barentin1999,barentin2000,elfring2016}.
The boundary conditions on the 3D bulk velocity $\mathbf{u}(\mathbf{r},z)$ are the stick (non-slip) conditions at the bottom surface of the bulk fluid ($z=0$) and at its top surface ($z=h$):
\begin{align}
\mathbf{u}(\mathbf{r},0) = 0,~~~~~
\mathbf{u}(\mathbf{r},h) = \mathbf{v}(\mathbf{r}),
\label{eq:bc}
\end{align}
with $\mathbf{v}(\mathbf{r})$ being the in-plane velocity vector of the fluid layer at $z=h$.
We now use the lubrication approximation where the vertical component of the velocity, $u_z$, is neglected compared to the in-plane components and the vertical pressure gradient vanishes, i.e.,
$\partial p/\partial z=0$.
This assumption is justified as long as $h$ is smaller than any horizontal characteristic length scale of the bulk fluid velocity.
Under these assumptions, the 3D Stokes equation~(\ref{eq:bulkstokes}) reduces to
\begin{align}
\eta \frac{\partial^2}{\partial z^2}
\begin{pmatrix}
u_x\\
u_y
\end{pmatrix}
-\nabla p=0.
\end{align}

Taking into account the boundary conditions in Eq.~(\ref{eq:bc}), the above equation can be integrated to give
\begin{align}
\mathbf{u}(\mathbf{r}, z)=\frac{z^{2}-z h}{2 \eta} \nabla p(\mathbf{r})
+\frac{z}{h} \mathbf{v}(\mathbf{r}).
\label{eq:u}
\end{align}
Then, the force exerted on the 2D fluid layer by the bulk fluid beneath is calculated as~\cite{barentin1999,barentin2000,elfring2016}
\begin{align}
\mathbf{f}_{\rm b} = - \eta \left.\frac{\partial\mathbf{u}}{\partial z}\right|_{z=h}
= -\frac{h}{2}\nabla p - \frac{\eta}{h}\mathbf{v}.
\end{align}
Substituting the obtained $\mathbf{f}_{\rm b}$ into Eq.~(\ref{eq:balance}), one can show that the hydrodynamic equation for the active chiral layer is
\begin{align}
& -\nabla \Pi+\eta_{\rm d}\nabla (\nabla\cdot\mathbf{v}) + \eta_{\rm s}\nabla^2\mathbf{v} + \eta_{\rm o}\nabla^2 \mathbf{v}^\ast \nonumber \\
&-\frac{h}{2}\nabla p - \frac{\eta}{h}\mathbf{v} +\mathbf{F}= 0,
\label{eq:monolayer}
\end{align}
where the divergence of the in-plane velocity is given by the following relation~\cite{barentin1999,elfring2016}
\begin{align}
\nabla\cdot\mathbf{v} = \frac{h^2}{6\eta}\nabla^2 p.
\label{eq:compress}
\end{align}
This relation can be derived by taking the divergence of Eq.~(\ref{eq:u}) and integrating over the lubrication layer (${0\leq z \leq h}$) under the incompressibility condition of Eq.~(\ref{eq:bulkincomp}) and the boundary conditions of Eq.~(\ref{eq:bc}).
The fourth term on the left-hand-side of Eq.~(\ref{eq:monolayer}) indicates that odd viscosity contributes to the fluid flow that is perpendicular to the one generated by shear viscosity~\cite{banerjee2017}.

\section{Hydrodynamic response of a point force}
\label{sec:mobility}

\subsection{Mobility tensor}

We derive the mobility tensor for the hydrodynamic response of the 2D fluid layer with odd viscosity.
The second-rank mobility tensor $\mathbf{G}(\mathbf{r})$ connects the force density $\mathbf{F}$ acting
on the fluid layer at position $\mathbf{r}^\prime$ with its induced velocity at position $\mathbf{r}$:
\begin{align}
v_i(\mathbf{r}) = \int {\rm d}^2 r^\prime\, G_{ij}(\mathbf{r}-\mathbf{r}^\prime) F_j(\mathbf{r}^\prime).
\label{eq:velocity}
\end{align}
In order to derive $G_{ij}(\mathbf{r})$, we solve the hydrodynamic equations~(\ref{eq:monolayer}) and (\ref{eq:compress})
in Fourier space and obtain $G_{ij}[\mathbf{k}]$, where
$\mathbf{k}=(k_x,k_y)$, and the square brackets indicate a function in Fourier space.
We introduce two orthogonal unit vectors
\begin{align}
\hat{\mathbf{k}}=(k_x/k,k_y/k),~~~~~
\bar{\mathbf{k}}=(-k_y/k,k_x/k),
\label{unitkvectors}
\end{align}
with $k=\vert \mathbf{k} \vert$.

\begin{figure*}[thb]
\centering\includegraphics[scale=0.38]{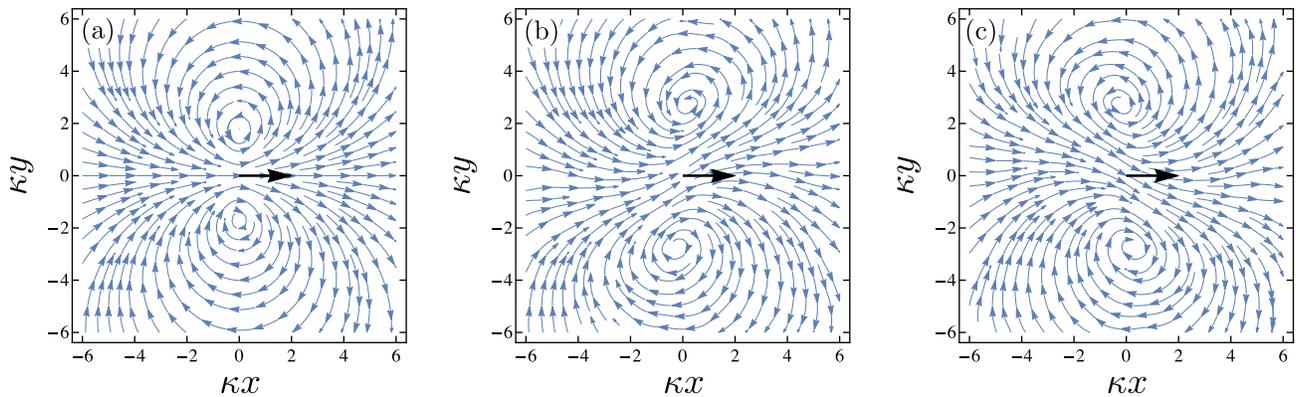}
\caption{
Streamlines of the velocity $\mathbf{v}(x,y)$ rescaled by $F/(2\pi\eta_{\rm s})$ and generated by a point force, $\mathbf{F}=F\hat{\mathbf{e}}_x\delta({\mathbf{r}})$, as a function of $\kappa x$ and $\kappa y$ while keeping $\eta_{\rm d}=3\eta_{\rm s}$.
The force along the $x$-axis is applied at the origin (the black horizontal arrow) for (a) $\mu=\eta_{\rm o}/\eta_{\rm s}=0$, (b) $\mu=3$, and (c) $\mu=-3$ [see Eqs.~(\ref{eq:velocity}) and (\ref{eq:realGij})].
The blue arrows indicate the flow direction.
\label{fig:v_x_y_mu}
}
\end{figure*}

\begin{figure*}[tbh]
\begin{center}
\includegraphics[scale=0.38]{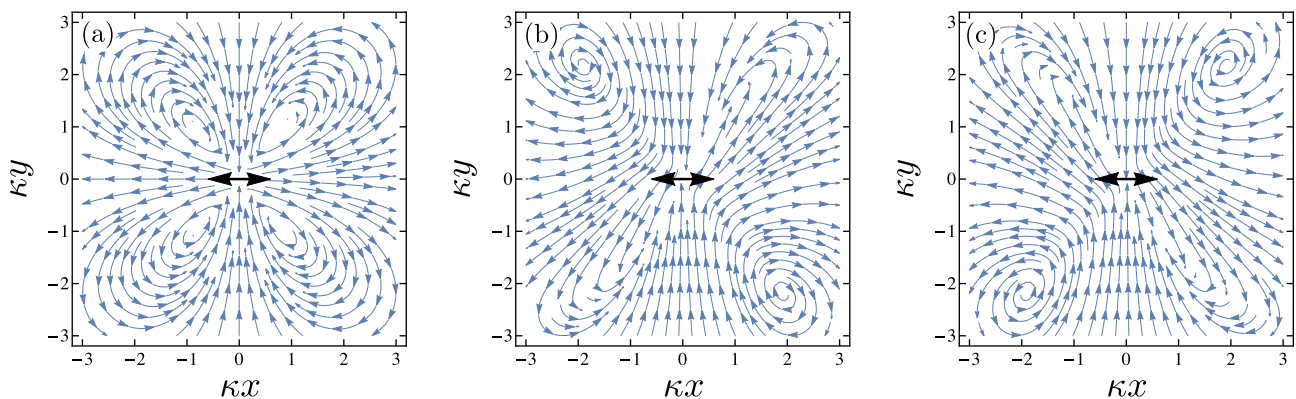}
\end{center}
\caption{
Streamlines of the velocity $\mathbf{v}(x,y)$ rescaled by $F\kappa\ell/(2\pi\eta_{\rm s})$ and generated by a force dipole as a function of $\kappa x$ and $\kappa y$ while keeping $\eta_{\rm d}=3\eta_{\rm s}$.
The force dipole along the $x$-axis ($\hat{\mathbf{d}}=\hat{\mathbf{e}}_x$) is centered at the origin (the black double arrow) for (a) $\mu=\eta_{\rm o}/\eta_{\rm s}=0$, (b) $\mu=3$, and (c) $\mu=-3$ [see Eq.~(\ref{eq:dipole})].
The blue arrows indicate the flow direction.
\label{fig:dipole}
}
\end{figure*}

In Appendix~\ref{app:Gijft}, we show that $G_{ij}[\mathbf{k}]$ has the following expression
\begin{align}
&G_{ij}[\mathbf{k}] \nonumber \\
&=
\frac{
\eta_{\rm s}(k^2+\kappa^2)\hat{k}_i\hat{k}_j
 +(\eta_{\rm s}+\eta_{\rm d})(k^2+\lambda^2)\bar{k}_i\bar{k}_j
-\eta_{\rm o}k^2\epsilon_{ij}
}
{\eta_{\rm s}(\eta_{\rm s}+\eta_{\rm d})(k^2+\kappa^2)(k^2+\lambda^2)+\eta_{\rm o}^2k^4},
\label{eq:Gij}
\end{align}
where
\begin{align}
\kappa^2=\frac{\eta}{h\eta_{\rm s}},~~~~~
\lambda^2=\frac{4\eta}{h(\eta_{\rm s}+\eta_{\rm d})}.
\label{kappalambda}
\end{align}
Note that the ratio, $\eta/\eta_{\rm s}$, gives the inverse length scale because the 2D viscosity, $\eta_{\rm s}$, has the dimension of Pa$\cdot$s$\cdot$m, while that of $\eta$ is Pa$\cdot$s.
The lengths scales, $\kappa^{-1}$ and $\lambda^{-1}$, correspond to the hydrodynamic screening lengths beyond which the 2D layer exchanges momentum with the underlying bulk fluid.
Importantly, the numerator of Eq.~(\ref{eq:Gij}) includes an anti-symmetric tensor $\epsilon_{ij}$.

\begin{figure*}[tbh]
\begin{center}
\includegraphics[scale=0.3]{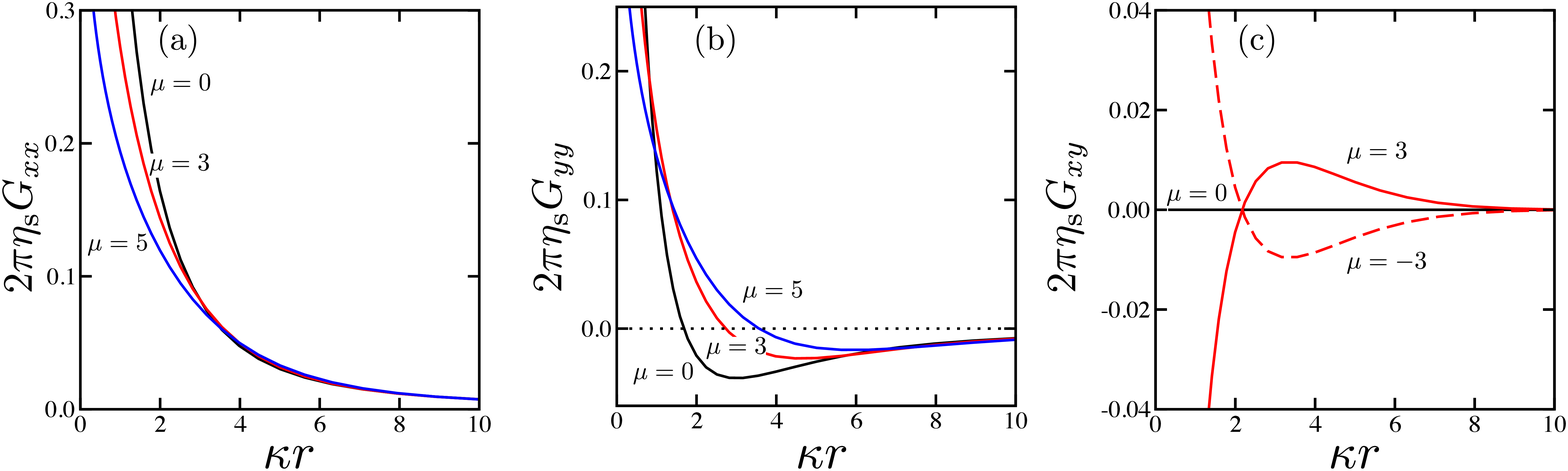}
\end{center}
\caption{
Plots of the mobility coefficients rescaled by $2\pi\eta_{\rm s}$ as a function of $\kappa r$ for various values of $\mu$.
(a) The longitudinal mobility coefficient $G_{xx}$ for $\mu=0$, $3$, and $5$ (black, red, and blue lines).
(b) The transverse mobility coefficient $G_{yy}$ for $\mu=0$, $3$, and $5$ (black, red, and blue lines).
(c) The anti-symmetric mobility coefficient $G_{xy}$ for $\mu=-3$, $0$, and $3$ (red dashed, black solid, and red solid lines).
\label{fig:G}
}
\end{figure*}

For simplicity sake, we hereafter assume $\kappa=\lambda$ (or equivalently
$\eta_{\rm d}=3\eta_{\rm s}$) in Eq.~(\ref{eq:Gij}) and consider the following simplified mobility tensor
\begin{align}
G_{ij}[\mathbf{k}] &=
\frac{(k^2+\kappa^2)(4\delta_{ij}-3\hat{k}_i\hat{k}_j) - \mu k^2\epsilon_{ij}}
{\eta_{\rm s}\left[4(k^2+\kappa^2)^2+\mu^2k^4\right]},
\label{eq:simpleGij}
\end{align}
where 
\begin{align}
\mu=\frac{\eta_{\rm o}}{\eta_{\rm s}}.
\end{align}
The above dimensionless parameter, $\mu$, is a
measure of how far the 2D active chiral fluid departs from the passive fluid, e.g., due to the self-spinning active objects.

As shown in Appendix~\ref{app:Gijreal}, the real space representation of the mobility tensor can be obtained by the inverse Fourier transform of Eq.~(\ref{eq:simpleGij})~\cite{doi1988}
\begin{align}
G_{ij}(\mathbf{r})= C_1(r)\delta_{ij}+C_2(r)\hat{r}_i\hat{r}_j + C_3(r)\epsilon_{ij},
\label{eq:realGij}
\end{align}
where $\hat{\mathbf{r}}=\mathbf{r}/r$ is a unit vector ($r=|\mathbf{r}|$), and the three coefficients
are given by
\begin{align}
C_1(r) & = \frac{1}{2\pi\eta_{\rm s}}
\int_0^\infty {\rm d}k\,
\frac{k(k^2+\kappa^2)}{4(k^2+\kappa^2)^2+\mu^2k^4}
\nonumber \\
&\times \left[ 4J_0(kr) - \frac{3J_1(kr)}{kr} \right],
\label{eq:c1}
\end{align}
\begin{align}
C_2(r) & =\frac{1}{2\pi\eta_{\rm s}}
\int_0^\infty {\rm d}k\,
\frac{3k(k^2+\kappa^2)}{4(k^2+\kappa^2)^2+\mu^2k^4}
\nonumber \\
& \times \left[ -J_0(kr) + \frac{2J_1(kr)}{kr} \right],
\label{eq:c2}
\end{align}
\begin{align}
C_3(r) =  -\frac{\mu}{2\pi\eta_{\rm s}}
\int_0^\infty {\rm d}k \,
\frac{k^3J_0(kr)}{4(k^2+\kappa^2)^2+\mu^2k^4}.
\label{eq:c3}
\end{align}
In the above, $J_n(x)$ is the Bessel function of the first kind~\cite{abramowitzhandbook}.
When $\eta_{\rm o}=0$ (or $\mu=0$), $C_3$ vanishes and $G_{ij}(\mathbf{r})$ reduces to that of a 2D passive compressible fluid, $G_{ij}^0(\mathbf{r})$, which we analytically derive in Appendix~\ref{app:Gijreal}
[see Eq.~(\ref{eq:Gij0})].
Hereafter, the superscript $``0"$ denotes quantities when $\mu=0$ (vanishing odd viscosity).
Under the exchange $\eta_{\rm o}\leftrightarrow -\eta_{\rm o}$,
$C_1$ and $C_2$ of Eqs.~(\ref{eq:c1}) and (\ref{eq:c2}) remain unchanged, whereas $C_3$ of Eq.~(\ref{eq:c3}) changes its sign.

We briefly discuss the symmetry property of the mobility tensor obtained in Eq.~(\ref{eq:realGij}).
From $G_{ij}^0$ in Eq.~(\ref{eq:Gij0}), one sees that the mobility tensor of passive fluids satisfies the symmetry property $G^0_{ij}=G^0_{ji}$, whereas for active chiral fluids, such a symmetry does not hold, i.e., $G_{ij}\neq G_{ji}$ as shown in Eq.~(\ref{eq:realGij}).
This asymmetry gives rise to the fluid velocity perpendicular to an applied force that results from the non-reciprocal hydrodynamic response.

\subsection{Velocity field}

With the obtained mobility tensor, we first investigate the velocity field induced by a point force acting on a
2D fluid layer with odd viscosity.
Substituting Eq.~(\ref{eq:realGij}) into Eq.~(\ref{eq:velocity}), we calculate the velocity field induced by a point force at the origin, $\mathbf{F}=F\hat{\mathbf{e}}_x\delta({\mathbf{r}})$, with $\hat{\mathbf{e}}_x$ being a unit vector in the $x$-direction.
The obtained velocity field is plotted in Fig.~\ref{fig:v_x_y_mu} for $\mu=0$, $3$, and $-3$.
When $\mu=0$, we see axisymmetric streamlines that pass through the applied force (the black horizontal arrow),
as in Fig.~\ref{fig:v_x_y_mu}(a).
There are two vortices whose center is located at $(\kappa x,\kappa y)\approx(0,\pm2.0)$.
They result from the nature of the 2D fluid layer with the hydrodynamic screening length, $\kappa^{-1}$.

When $\mu$ is finite, however, a perpendicular flow in the $y$-direction starts to develop and accordingly, the axial 
symmetry breaks down, as shown in Figs.~\ref{fig:v_x_y_mu}(b) and (c).
This behavior results from the non-reciprocal hydrodynamic response in the presence of the odd viscosity.
When $\mu=\pm 3$, the vortices approach to the positions $(0,\pm3.0)$, meaning that finite values of
$\mu$ causes an effective increase in the hydrodynamic length, $\kappa^{-1}$.

We next calculate the flow field generated by a hydrodynamic force dipole that is composed of two point forces directed oppositely to each other.
When a force dipole at the origin is directed along a given unit vector $\hat{\mathbf{d}}$, its induced velocity field is given by~\cite{spagnolie2012}
\begin{align}
v_i(\mathbf{r}) = -F\ell \hat{d}_k \partial_k G_{ij}(\mathbf{r})\hat{d}_j.
\label{eq:dipole}
\end{align}
Here, $F$ is the force magnitude, $\ell$ is the distance between the two point forces, and the limit $\ell\ll r$ is assumed.
In Fig.~\ref{fig:dipole}, we plot the force dipole along the $x$-axis (i.e., $\hat{\mathbf{d}}=\hat{\mathbf{e}}_x$) for $\mu=0$, $3$, and $-3$.
When $\mu=0$, the streamlines have an axial symmetry along the $x$-direction as well as the $y$-direction, as shown in Fig.~\ref{fig:dipole}(a).
For $\mu=\pm 3$, however, flows perpendicular to the applied forces become dominant, and both mirror symmetries are broken, as shown in Fig.~\ref{fig:dipole}(b) and (c).

\subsection{Mobility coefficients}

\begin{figure}[tbh]
\begin{center}
\includegraphics[scale=0.35]{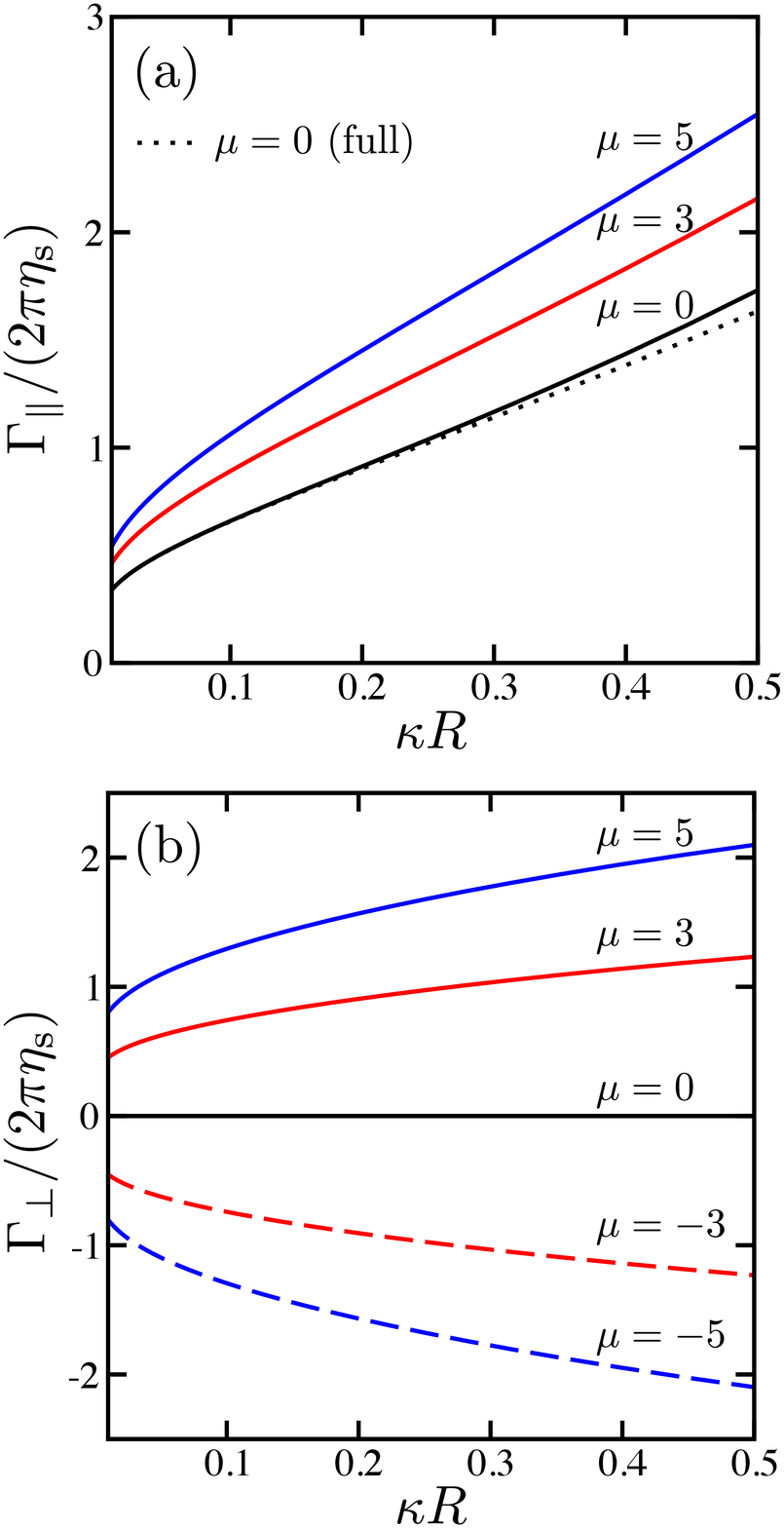}
\end{center}
\caption{
Plots of the drag ($\Gamma_\|$) and lift ($\Gamma_\perp$) coefficients rescaled by $2\pi\eta_{\rm s}$ as a function of the rescaled disk radius $\kappa R$.
(a) The drag coefficient $\Gamma_\|$ for $\mu=0$, $3,$ and $5$ (black, red, and blue solid lines).
The dotted line represents the full expression $\Gamma_\|^0$
for the drag coefficient when $\mu=0$ reported in Ref.~\cite{barentin1999} (see the text for the specific expression).
(b) The lift coefficient $\Gamma_\perp$ for $\mu=-5$, $-3$, $0$, $3$, and $5$ (blue dashed, red dashed, black solid, red solid, and blue solid lines).
}
\label{fig:K}
\end{figure}

\begin{figure}[tbh]
\begin{center}
\includegraphics[scale=0.35]{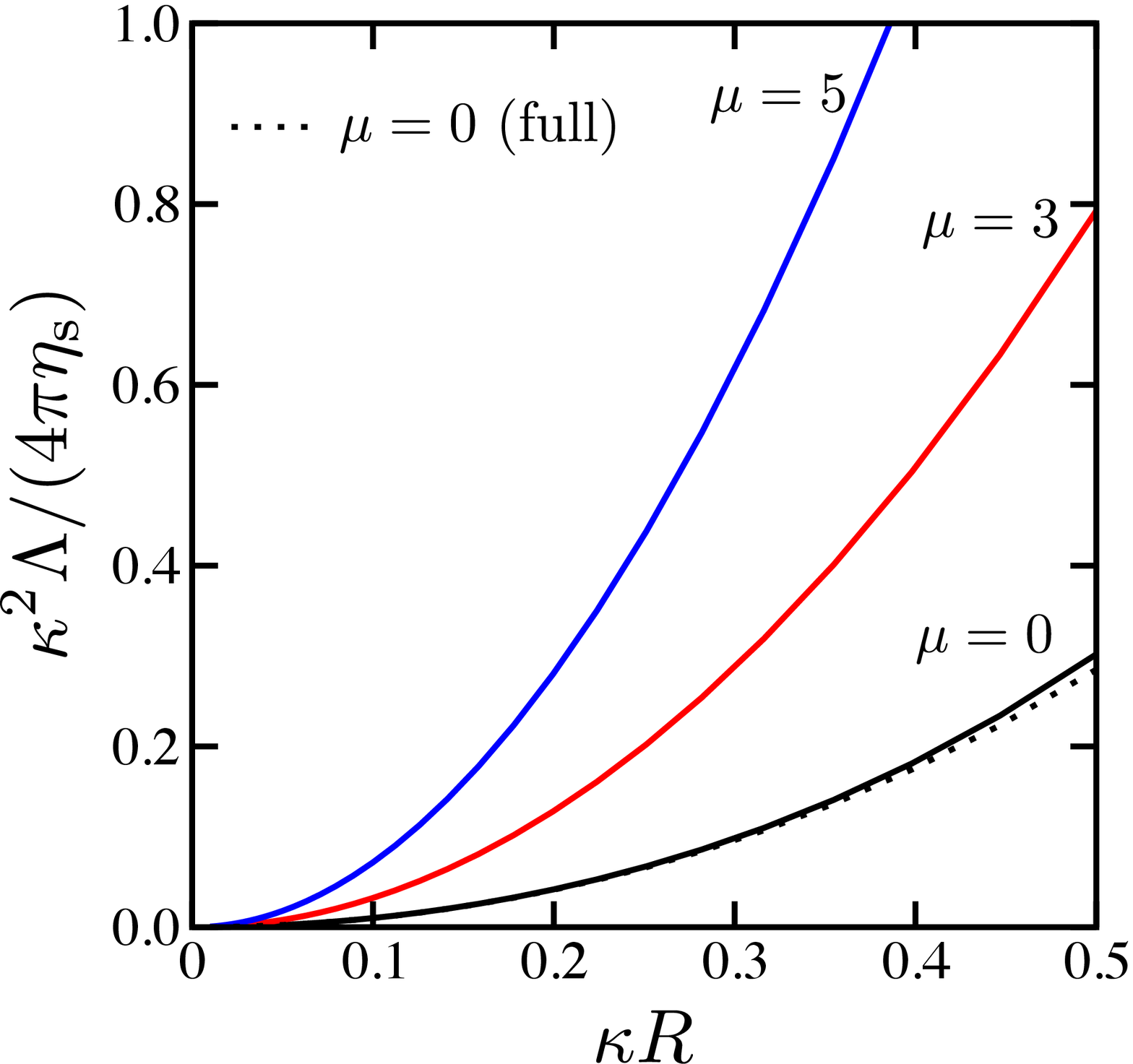}
\end{center}
\caption{
Plot of the rotational friction coefficient ($\Lambda$) rescaled by $4\pi\eta_{\rm s}/\kappa^2$
as a function of the rescaled disk radius $\kappa R$ for $\mu=0$, $3$, and $5$ (black, red, and blue solid lines).
The dotted line represents the full expression $\Lambda^0$ 
for the rotational coefficient when $\mu=0$ reported in Ref.~\cite{evans1988} (see the text for the specific expression).
}
\label{fig:L}
\end{figure}

Next, we investigate each component of the mobility tensor in Eq.~(\ref{eq:realGij}).
If we choose the $x$-axis, without loss of generality, along the $r$ direction, i.e., $\mathbf{r}=r\hat{\mathbf{e}}_x$, the longitudinal, transverse, and anti-symmetric mobility coefficients are given by
$G_{xx}=C_1+C_2$, $G_{yy}=C_1$, and $G_{xy}=-G_{yx}=C_3$, respectively.
The non-zero mobility coefficient $G_{xy}$ is characteristic of the active chiral fluid with finite odd viscosity, while $G_{xx}$ and $G_{yy}$ remain in the limit of $\eta_{\rm o}\to0$.
Note that both $G_{xx}$ and $G_{yy}$ also depend on $\eta_{\rm o}$, as seen in Eqs.~(\ref{eq:c1}) and (\ref{eq:c2}).

In Fig.~\ref{fig:G}, we plot $G_{xx}$, $G_{yy}$, and $G_{xy}$ as a function of $\kappa r$ for various values
of $\mu$, while keeping $\eta_{\rm d}=3\eta_{\rm s}$.
We see that $G_{xx}$ decreases monotonically with $\kappa r$ for all the $\mu$ values, as shown in Fig.~\ref{fig:G}(a).
The decrease of $G_{xx}$ is more enhanced for lager $\mu$, whereas $G_{xx}$ weakly depends on $\mu$ for
larger $\kappa r$.
This reflects the fact that, for $\kappa r\gg1$, the 3D hydrodynamic effect becomes more important, and $G_{xx}$
is almost independent of the odd viscosity $\eta_{\rm o}$.
In Fig.~\ref{fig:G}(b), on the other hand, $G_{yy}$ takes negative values because the 2D fluid layer can flow in the direction opposite to the applied force~\cite{oppenheimer2010}.
Such a behavior results from the two vortices shown in Fig.~\ref{fig:v_x_y_mu}.

The mobility coefficient $G_{xy}$ describes the non-reciprocal hydrodynamic response because it
gives the relation between the applied force $F_y$ and induced velocity $v_x$.
When $\mu=0$, $G_{xy}$ is always zero as it should be, whereas for $\mu=\pm 3$, it exhibits
non-monotonic behavior, as shown in Fig.~\ref{fig:G}(c).
This means that when the system is active, an applied force in the 2D fluid layer generates a
perpendicular flow, giving rise to the broken mirror symmetry with respect to the force direction,
even when the fluid layer is isotropic.
When $\eta_{\rm o}$ is positive, $G_{xy}$ takes negative ($\kappa r<2$) and positive values ($\kappa r>2$), corresponding to the attractive and repulsive flows, respectively, and vice versa for negative $\eta_{\rm o}$.
These flow patterns can lead to either convergence or dispersion of surrounding inclusions, and the specific behavior depends on the hydrodynamic screening length, $\kappa^{-1}$.

\section{Hydrodynamic response of a rigid disk}
\label{sec:force}

\subsection{Boundary integral equation}

So far, we have discussed the hydrodynamic response induced by a point force and a force dipole in a 2D fluid layer with odd viscosity using the mobility tensor in Eq.~(\ref{eq:realGij}).
Here we generalize the discussion to the situation where the response is induced by a finite-size body moving in the 2D fluid layer.
For a passive fluid, the force acting on such a body can be calculated by using a boundary integral equation that is 
based on the Lorentz reciprocal theorem~\cite{pozrikidis1992,masoud2019}.
In Appendix~\ref{app:nreciprocal}, we first generalize this theorem for a 2D compressible fluid with finite $\eta_{\rm o}$.
Then, in Appendix~\ref{app:BIE}, we derive the corresponding boundary integral equation that is used in the following analysis.

Consider a circular rigid disk of radius $R$, which translates and rotates in the 2D fluid.
We assume that a no-slip boundary condition holds at the disk perimeter and further consider the limit of $\kappa R\ll1$.
As detailed in Appendix~\ref{app:BIE}, the velocity at any point on the disk perimeter ($\mathbf{R}=R\hat{\mathbf{r}}$) 
can be expressed in terms of the following boundary integral equation
\begin{align}
U_{i}+\epsilon_{ijk}\Omega_jR_k=-\int_{C_{\rm u}} {\rm d}s(R^\prime)\,
f_{j}\left(\mathbf{R}^{\prime}\right)
G_{ji}\left(\mathbf{R}-\mathbf{R}^{\prime}\right),
\label{eq:noslip}
\end{align}
where $\mathbf{U}$ and $\mathbf{\Omega}$ are the lateral and angular velocities of the rigid disk, respectively, and $\epsilon_{ijk}$ is the 3D Levi-Civita tensor.
The right-hand-side of Eq.~(\ref{eq:noslip}) is a line integral over an unspecified closed curve $C_{\rm u}$, and ${\rm d}s(R^\prime)$ indicates that $\mathbf{R}^\prime$ is the integration variable.
The boundary integral equation~(\ref{eq:noslip}) relates the velocities of the disk moving in the 2D fluid layer with the accompanying unknown force distribution $\mathbf{f}$.

\subsection{Translational and rotational frictions}

Since the governing hydrodynamic equation~(\ref{eq:monolayer}) is linear in $\mathbf{v}$,
the translational motion is decoupled from the rotational one.
Hence, the following linear relations hold
\begin{align}
\mathbf{F}^{\rm d} = -\mathbf{\Gamma}\cdot\mathbf{U},~~~~~
\mathbf{T}^{\rm d} = -\Lambda \mathbf{\Omega},
\label{eq:KL}
\end{align}
where $F^{\rm d}_i=\int_{C_{\rm u}} {\rm d}s(R^\prime)\, f_i(\mathbf{R}^\prime)$ is the force acting on the disk and $T^{\rm d}_i=\epsilon_{ijk}\int_{C_{\rm u}} {\rm d}s(R^\prime)\, R_j^\prime f_k(\mathbf{R}^\prime)$ is the torque, while $\mathbf{\Gamma}$ is the translational friction tensor and $\Lambda$ is the rotational friction coefficient.
The minus signs in Eq.~(\ref{eq:KL}) take into account that the force and torque act opposite to the velocities.
Note that $\Lambda$ is a scalar because both $\mathbf{T}^{\rm d}$ and $\mathbf{\Omega}$ must point
to the $z$-direction in a 2D system.

Using the assumption $\vert \mathbf{R}^\prime \vert \ll \vert \mathbf{R} \vert$ in Eq.~(\ref{eq:noslip})~\cite{oppenheimer2009},
we obtain the expressions for $\mathbf{\Gamma}$ and $\Lambda$ as 
\begin{align}
\mathbf{\Gamma} =
\frac{1}{\left(C_1+C_2/2\right)^2+C_3^2}
\begin{pmatrix}
C_1+C_2/2 & C_3 \\
-C_3 & C_1+C_2/2  \\
\end{pmatrix},
\label{eq:translation}
\end{align}
\begin{align}
\Lambda&=\frac{2R^2}{C_2 - R(\partial C_1/\partial R)}.
\label{eq:rotation}
\end{align}
See Appendix~\ref{app:resistance} for the derivation.
In the above, the arguments of the three coefficients are omitted, $C_n\equiv C_n(R)~(n=1,2,3)$, in order to keep the notation compact.
For passive fluids, the translational friction tensor must be symmetric and positive definite
according to the requirement that the dissipated energy is positive~\cite{happel2012}.
For the considered fluid with $\eta_{\rm o}$, however, the translational friction tensor is allowed to be asymmetric 
when $C_3$ is nonzero.
Notice that the energy dissipation calculated from Eq.~(\ref{eq:translation}) is $U_i\Gamma_{ij}U_j\sim(C_1+C_2/2)U^2$ and the anti-symmetric part does not contribute to the dissipation.

For a disk translating with the velocity $\mathbf{U}=(U,0)$,
we have the viscous drag and lift coefficients as $\Gamma_\|=\Gamma_{xx}$ and
$\Gamma_\perp=\Gamma_{yx}$, respectively.
In Fig.~\ref{fig:K}, we plot $\Gamma_\|$ and $\Gamma_\perp$ as a function of the dimensionless radius $\kappa R$ for various values of $\mu$.
We see that $\Gamma_\|$ increases monotonically with increasing the disk size for all $\mu$ values, as seen in Fig.~\ref{fig:K}(a).
For a fixed disk size, $\Gamma_\|$ is larger for larger $\mu$ values.
The dotted line in Fig.~\ref{fig:K}(a) is the full analytical result for $\mu=0$ (2D passive compressible fluid)~\cite{barentin1999},
$\Gamma^0_\|/(2\pi\eta_{\rm s})=\frac{4}{5}(\kappa R)^2K_2(\kappa R)/K_0(\kappa R)$,
where $K_n(x)$ is the modified Bessel function of the second kind~\cite{abramowitzhandbook}.
Our result obtained by using the boundary integral equation~(\ref{eq:noslip}) coincides with the analytical one when $\kappa R\ll1$.

In Fig.~\ref{fig:K}(b), we see that $\Gamma_\perp$ shows both increasing and decreasing dependencies on
$\kappa R$ for positive and negative values of $\mu$, respectively.
However, when $\mu=0$, $\Gamma_\perp$ is always zero, as it should be.
Finite values of $\Gamma_\perp$ mean that the disk translated along the $x$-axis presents a lift motion along the $y$-direction.
For a fixed disk size, the absolute value of $\Gamma_\perp$ increases and the lift force is more enhanced
as the absolute value of $\mu$ increases.

Assuming that the disk is rotating with velocity $\mathbf{\Omega}=(0,0,\Omega)$, we obtain the
rotational friction coefficient $\Lambda$ that is plotted in Fig.~\ref{fig:L} as a function of $\kappa R$.
We see that $\Lambda$ shows an increasing dependency on $\kappa R$ for all the $\mu$ values.
The dotted line in Fig.~\ref{fig:L} represents the full analytical expression for $\mu=0$ (2D passive incompressible fluid)~\cite{evans1988},
$\kappa^2\Lambda^0/(4\pi\eta_{\rm s})=(\kappa R)^2+\frac{1}{2}(\kappa R)^3K_0(\kappa R)/K_1(\kappa R)$.
The solid black line and the dotted line coincide in the limit of $\kappa R\ll1$, because the
disk rotation contributes neither to the compression nor to the expansion of fluids.

\section{Discussion and conclusion}
\label{sec:discussion}

We have investigated the linear hydrodynamic response of a 2D fluid layer with broken time-reversal and parity symmetries.
Such a 2D active fluid presents a special rheological property called \textit{odd viscosity}, characterizing the deviation of the system from a passive fluid.
In our approach, we combine the concept of the odd viscosity~\cite{avron1998,banerjee2017,epstein2020} and the hydrodynamic model of a 2D compressible fluid derived by using the lubrication 
approximation~\cite{barentin1999,barentin2000,elfring2016}.
In contrast to well-studied 2D fluids characterized by a shear viscosity~\cite{evans1988,diamant2009,oppenheimer2009,oppenheimer2010,ramachandran2011,ramachandran2010}, the additional odd viscosity $\eta_{\rm o}$ leads to anomalous flow behavior, i.e., non-reciprocal hydrodynamic response.

In the case of a point force and a force dipole, the symmetry of the velocity field in terms of the force direction is broken, generating flow perpendicular to the applied force (see Figs.~\ref{fig:v_x_y_mu} and \ref{fig:dipole}).
We also analyzed the effects of the odd viscosity on the mobility tensor, as derived in Eq.~(\ref{eq:realGij}).
In particular, we investigated the behavior of the anti-symmetric mobility coefficient $G_{xy}$ that exists only for non-vanishing $\eta_{\rm o}$, as shown in Eq.~(\ref{eq:c3}).
As for the hydrodynamic response of finite-size bodies, we have investigated the forces acting on a translating and rotating disk in a 2D fluid layer, using the boundary integral equation~(\ref{eq:noslip}).
We found that small disks ($\kappa R\ll1$) not only undergo a drag force, but also a lift force, which
cannot be seen in an isotropic passive fluid (see Fig.~\ref{fig:K})~\cite{avron1998}.

As a possible biological application, we can relate the 2D fluid layer to a monolayer with a low concentration of active motor proteins, such as ion pumps~\cite{albertsbook}.
Rheological properties of these system can be investigated by surface microrheology techniques~\cite{maestro2011}.
For typical values such as
$\eta\approx10^{-3}$\,Pa$\cdot$s,
$\eta_{\rm s}\approx10^{-6}$\,Pa$\cdot$s$\cdot$m,
and $h\approx1$\,nm,
we find that the obtained drag and lift forces could be observed in experiments using a sub-micrometer probe, i.e., $R<0.1$\,$\mu$m.

In this paper, we have considered a 2D compressible fluid, which is in contact with a 3D bulk fluid.
Such a compressible 2D system can be realized by a dilute Gibbs monolayer, which is composed of soluble amphiphiles that can dissolve into the underlying bulk fluid~\cite{barentin1999,barentin2000}.
When the adsorption and desorption processes of soluble amphiphiles are instantaneous, surface concentration gradients can be eliminated rapidly.
This is the situation that we consider in the present work.
When the amphiphile is insoluble, on the other hand, the concentration gradient is sustained, giving rise to a Marangoni flow~\cite{elfring2016,manikantan2020}.
Although the Marangoni convection is outside the scope of this paper, it would be interesting to investigate the effects of odd viscosity on such convective phenomena.

In general, odd viscosity can depend on the density of self-spinning objects, although such an effect was not considered in the present work.
By using the 2D Fax\'{e}n laws, effective shear viscosity of a fluid membrane with finite-size suspensions was derived~\cite{oppenheimer2009,oppenheimer2010}.
Hence, it is of interest to see how the non-reciprocal flow field of the 2D fluid with an odd viscosity $\eta_{\rm o}$ changes with the rotor concentration.
Moreover, for a passive fluid at equilibrium, the disk drag coefficient is connected to its diffusion constant through Einstein's relation.
However, such an evident relation does not exist in active chiral fluids and one has to extend the fluctuation dissipation theorem~\cite{yasuda2017,hosaka2017,yasuda2017_2,hosaka2020_2}.
These interesting questions are left for future investigations.

\acknowledgements

We thank J.\ E.\ Avron, H.\ Diamant, M.\ Doi, R.\ Seto, M.\ S.\ Turner, and K.\ Yasuda for useful discussions.
Y.H.\ acknowledges support by a Grant-in-Aid for JSPS Fellows (Grant No.\ 19J20271) from the Japan Society for the Promotion of Science (JSPS).
Y.H.\ also thanks the hospitality of Tel Aviv University, where part of this research was conducted under the TMU-TAU co-tutorial program.
S.K.\ acknowledges support by a Grant-in-Aid for Scientific Research (C) (Grant No.\ 18K03567 and Grant No.\ 19K03765) from the JSPS, and support by a Grant-in-Aid for Scientific Research on Innovative Areas ``Information physics of living matters'' (Grant No.\ 20H05538) from the Ministry of Education, Culture, Sports, Science and Technology of Japan.
D.A.\ acknowledges support from the Israel Science Foundation (ISF) under Grant No.\ 213/19.

\begin{widetext}

\appendix
\section{Derivation of Eq.~(\ref{eq:Gij})}
\label{app:Gijft}

We derive the mobility tensor in Fourier space $\mathbf{G}[\mathbf{k}]$ as given by  Eq.~(\ref{eq:Gij}).
The Fourier transform of $\mathbf{v}(\mathbf{r})$ is defined by
\begin{align}
\mathbf{v}(\mathbf{r})=\int \frac{{\rm d}^{2} k}{(2 \pi)^{2}} \, \mathbf{v}[\mathbf{k}] \exp (i \mathbf{k} \cdot \mathbf{r}),
\end{align}
with $\mathbf{k}=(k_x,k_y)$, and similarly for the 3D pressure $p(\mathbf{r})$
and the force density $\mathbf{F}(\mathbf{r})$.
In the Fourier space, Eq.~(\ref{eq:monolayer}) becomes
\begin{align}
 -\eta_{\rm s}k^2\mathbf{v}[\mathbf{k}]
-\eta_{\rm d} k^2 \hat{\mathbf{k}} \hat{\mathbf{k}}\cdot\mathbf{v}[\mathbf{k}]
-\eta_{\rm o} k^2 (\hat{\mathbf{k}}\bar{\mathbf{k}}\cdot\mathbf{v}[\mathbf{k}]-\bar{\mathbf{k}}\hat{\mathbf{k}}\cdot\mathbf{v}[\mathbf{k}]) 
 -\frac{ih}{2} k p[\mathbf{k}] \hat{\mathbf{k}}
-\frac{\eta}{h}\mathbf{v}[\mathbf{k}]  + \mathbf{F}[\mathbf{k}] = 0,
\label{eq:monolayerft}
\end{align}
or equivalently
\begin{align}
 -\eta_{\rm s}k^2\mathbf{v}[\mathbf{k}]
-\eta_{\rm d} k^2 v_\|[\mathbf{k}]\hat{\mathbf{k}} 
-\eta_{\rm o} k^2 (v_\perp[\mathbf{k}]\hat{\mathbf{k}}-v_\|[\mathbf{k}]\bar{\mathbf{k}}) 
-\frac{ih}{2} k p[\mathbf{k}] \hat{\mathbf{k}}
-\frac{\eta}{h}\mathbf{v}[\mathbf{k}]  + \mathbf{F}[\mathbf{k}] = 0, 
\end{align}
where $v_\|[\mathbf{k}]=\hat{\mathbf{k}}\cdot\mathbf{v}[\mathbf{k}]$ and $v_\perp[\mathbf{k}]=\bar{\mathbf{k}}\cdot\mathbf{v}[\mathbf{k}]$.
For Eq.~(\ref{eq:compress}), we have
\begin{align}
ik\hat{\mathbf{k}}\cdot\mathbf{v}[\mathbf{k}] 
=ikv_\|[\mathbf{k}]
= -\frac{h^2}{6\eta}k^2p[\mathbf{k}].
\label{eq:compressft}
\end{align}
In the derivation of Eq.~(\ref{eq:monolayerft}), we have assumed that the 2D fluid layer is quickly equilibrated with the 3D bulk, and hence the 2D pressure is homogeneous in space, i.e., $\nabla \Pi=0$~\cite{barentin1999,elfring2016}.

Substituting Eq.~(\ref{eq:compressft}) into Eq.~(\ref{eq:monolayerft}) to eliminate $p[\mathbf{k}]$,
we obtain
\begin{align}
-\eta_{\rm s}k^2\mathbf{v}[\mathbf{k}]
-\eta_{\rm d} k^2 v_\|[\mathbf{k}]\hat{\mathbf{k}}
-\eta_{\rm o} k^2 (v_\perp[\mathbf{k}]\hat{\mathbf{k}}-v_\|[\mathbf{k}]\bar{\mathbf{k}})
-\frac{3\eta}{h} v_\|[\mathbf{k}] \hat{\mathbf{k}}
-\frac{\eta}{h}\mathbf{v}[\mathbf{k}]  + \mathbf{F}[\mathbf{k}]= 0.
\end{align}
Hence, $\mathbf{F}[\mathbf{k}]$ can be written as
\begin{align}
\displaystyle
	\begin{pmatrix}
		F_\|[\mathbf{k}] \\
		F_\perp[\mathbf{k}]
	\end{pmatrix}
	=
	\begin{pmatrix}
	(\eta_{\rm s}+\eta_{\rm d})k^2 + 4\eta/h & \eta_{\rm o}k^2 \\
	-\eta_{\rm o}k^2  & \eta_{\rm s}k^2 + \eta/h
	\end{pmatrix}
	\begin{pmatrix}
		v_\|[\mathbf{k}] \\
		v_\perp[\mathbf{k}]
	\end{pmatrix}.
\end{align}
Since the mobility tensor in the Fourier space satisfies the relation
$\mathbf{v}[\mathbf{k}]=\mathbf{G}[\mathbf{k}] \cdot \mathbf{F}[\mathbf{k}]$,
we obtain Eq.~(\ref{eq:Gij}).

\section{Derivation of Eq.~(\ref{eq:realGij}) and $\mathbf{G}^0(\mathbf{r})$}
\label{app:Gijreal}

Here we perform the inverse Fourier transform of $\mathbf{G}[\mathbf{k}]$ in Eq.~(\ref{eq:Gij})
to obtain $\mathbf{G}(\mathbf{r})$.
By calculating $G_{ii}$, $G_{ij}\hat{r}_i\hat{r}_j$, and $G_{ij}\epsilon_{ij}$, we obtain~\cite{doi1988}
\begin{align}
2C_1 + C_2 &= \int \frac{{\rm d}^2k}{(2\pi)^2}
\frac{\eta_{\rm s}(k^2+\kappa^2) +(\eta_{\rm s}+\eta_{\rm d})(k^2+\lambda^2)}
{\eta_{\rm s}(\eta_{\rm s}+\eta_{\rm d})(k^2+\kappa^2)(k^2+\lambda^2)+\eta_{\rm o}^2k^4}
\exp(i\mathbf{k}\cdot\mathbf{r}) \nonumber\\
&=
\frac{1}{2\pi\eta_{\rm s}}
\int_0^\infty {\rm d}k\,k
\frac{\eta_{\rm s}(k^2+\kappa^2)
+(\eta_{\rm s}+\eta_{\rm d})(k^2+\lambda^2)}
{(\eta_{\rm s}+\eta_{\rm d})(k^2+\kappa^2)(k^2+\lambda^2)+\eta_{\rm o}^2k^4}J_0(kr),
\label{eq:2c1c2}
\end{align}
\begin{align}
C_1+C_2&=
\int \frac{{\rm d}^2k}{(2\pi)^2}
\frac{ \eta_{\rm s}(k^2+\kappa^2)\cos^2\theta
+(\eta_{\rm s}+\eta_{\rm d})(k^2+\lambda^2)(1-\cos^2\theta)}
{\eta_{\rm s}(\eta_{\rm s}+\eta_{\rm d})(k^2+\kappa^2)(k^2+\lambda^2)+\eta_{\rm o}^2k^4}
\exp(i\mathbf{k}\cdot\mathbf{r}) \nonumber\\
&=
\frac{1}{2\pi\eta_{\rm s}}
\int_0^\infty {\rm d}k\,k
\frac{ \eta_{\rm s}(k^2+\kappa^2)[J_0(kr)-J_1(kr)/(kr)]
+(\eta_{\rm s}+\eta_{\rm d})(k^2+\lambda^2)J_1(kr)/(kr)}
{(\eta_{\rm s}+\eta_{\rm d})(k^2+\kappa^2)(k^2+\lambda^2)+\eta_{\rm o}^2k^4},
\label{eq:c1c2}
\end{align}
\begin{align}
 C_3 &= -\int \frac{{\rm d}^2k}{(2\pi)^2}
\frac{\eta_{\rm o} k^2}
{\eta_{\rm s}(\eta_{\rm s}+\eta_{\rm d})(k^2+\kappa^2)(k^2+\lambda^2)+\eta_{\rm o}^2k^4}
\exp(i\mathbf{k}\cdot\mathbf{r}) \nonumber\\
&=
-\frac{\eta_{\rm o}}{2\pi\eta_{\rm s}}\int_0^\infty {\rm d}k\,
\frac{k^3J_0(kr)}{(\eta_{\rm s}+\eta_{\rm d})(k^2+\kappa^2)(k^2+\lambda^2)+\eta_{\rm o}^2k^4},
\label{eq:appc3}
\end{align}
respectively, where $\theta$ is the angle between the vectors $\mathbf{k}$ and $\mathbf{r}$.
Solving Eqs.~(\ref{eq:2c1c2}) and (\ref{eq:c1c2}) when $\kappa=\lambda$,
we obtain Eqs.~(\ref{eq:c1}), (\ref{eq:c2}), and (\ref{eq:c3}).

Next, we analytically derive the mobility tensor $\mathbf{G}^0(\mathbf{r})$ in the absence
of the odd viscosity, i.e., $\mu=0$.
In this case, we have from Eq.~(\ref{eq:Gij})
\begin{align}
G^0_{ij}[\mathbf{k}] = \frac{\delta_{ij}-\hat{k}_i \hat{k}_j}{\eta_{\rm s}(k^2+ \kappa^2)}
+\frac{\hat{k}_i\hat{k}_j}{(\eta_{\rm s}+\eta_{\rm d})(k^2 + \lambda^2)}.
\end{align}
The real-space mobility tensor $\mathbf{G}^0(\mathbf{r})$ can be obtained by assuming
\begin{align}
G^0_{ij}(\mathbf{r})=B_{1}(r) \delta_{ij}+B_{2}(r) \hat{r}_i\hat{r}_j,
\label{eq:Gij0}
\end{align}
with two coefficients $B_1$ and $B_2$.

By calculating $G^0_{ii}$ and $G^0_{ij}\hat{r}_i\hat{r}_j$, we have
\begin{align}
2B_1+B_2 &= \frac{1}{2\pi} \int_0^\infty {\rm d}k\,
\left[ \frac{k}{\eta_{\rm s}(k^2+ \kappa^2)} 
+\frac{k}{(\eta_{\rm s}+\eta_{\rm d})(k^2 + \lambda^2)}
\right]J_0(kr) \nonumber\\
&= \frac{1}{2\pi\eta_{\rm s}}K_0(\kappa r) + \frac{1}{2\pi(\eta_{\rm s}+
\eta_{\rm d})}K_0(\lambda r),
\label{eq:2b1b2}
\end{align}
\begin{align}
B_1 + B_2 &=\frac{1}{2\pi} \int_0^\infty {\rm d}k\,
\left[ \frac{J_1(kr)}{\eta_{\rm s}r(k^2+ \kappa^2)}
+\frac{k}{(\eta_{\rm s}+\eta_{\rm d})(k^2 + \lambda^2)}
\left( J_0(kr) - \frac{J_1(kr)}{kr}\right) \right] \nonumber\\
&= \frac{1}{2 \pi \eta_{\rm s}}\left[-\frac{K_{1}(\kappa r)}{\kappa r}
+\frac{1}{(\kappa r)^2} \right]
+\frac{1}{2\pi(\eta_{\rm s}+\eta_{\rm d})}
\left[ K_0(\lambda r) + \frac{K_1(\lambda r)}{\lambda r} - \frac{1}{(\lambda r)^2}
\right].
\label{eq:b1b2}
\end{align}
Solving Eqs.~(\ref{eq:2b1b2}) and (\ref{eq:b1b2}), we obtain
\begin{align}
B_1(r) &= \frac{1}{2\pi\eta_{\rm s}} \left[ K_0(\kappa r)+\frac{K_1(\kappa r)}{\kappa r}
-\frac{1}{(\kappa r)^2}\right]
+\frac{1}{2\pi(\eta_{\rm s}+\eta_{\rm d})}\left[ -\frac{K_1(\lambda r)}{\lambda r}+\frac{1}{(\lambda r)^2}\right],
\label{eq:b1}
\end{align}
\begin{align}
B_2(r) = \frac{1}{2\pi\eta_{\rm s}}
\left[-K_{0}(\kappa r)-\frac{2 K_{1}(\kappa r)}{\kappa r}+\frac{2}{(\kappa r)^2}\right]
+\frac{1}{2\pi(\eta_{\rm s}+\eta_{\rm d})}
\left[ K_0(\lambda r)+ \frac{2K_1(\lambda r)}{\lambda r} -
\frac{2}{(\lambda r)^2} \right].
\label{eq:b2}
\end{align}

\section{Generalized Lorentz reciprocal theorem}
\label{app:nreciprocal}

The Lorentz reciprocal theorem gives a relation regarding the resistance of finite-size bodies moving in a passive fluid~\cite{happel2012,masoud2019}.
Here we generalize this theorem for a 2D compressible fluid with finite odd viscosity.
Let unprimed and primed symbols represent the variables for
any two arbitrary types of flows satisfying the following equations
\begin{align}
\partial_j \sigma_{ij} + b_i = 0,~~~~
\partial_j \sigma_{ij}^\prime + b_i^\prime = 0,
\label{eq:forcebalance}
\end{align}
where $\mathbf{b}$ and $\mathbf{b}^\prime$ are the arbitrary body force densities and the associated velocity fields are given by $\mathbf{v}$ and $\mathbf{v}^\prime$, respectively.
In the above, $\sigma_{ij}$ is the stress tensor in Eq.~(\ref{eq:stress}) that also includes the 2D isotropic pressure term, $-\Pi\delta_{ij}$ (similarly for $\sigma_{ij}^\prime$).
The divergence of $\sigma^\prime_{ij}v_{j}-\sigma_{ij}v_{j}^\prime$ becomes~\cite{cunha2003}
\begin{align}
\partial_i [(\sigma_{{\rm S}, ij}^\prime+\sigma_{{\rm A}, ij}^\prime) v_j] - \partial_i \left[(\sigma_{{\rm S}, ij}-\sigma_{{\rm A}, ij})v_j^\prime\right]&=v_j^\prime b_j - v_j b_j^\prime -\Pi^\prime\partial_jv_j+\Pi\partial_jv_j^\prime
+2 v_j^\prime \partial_i \sigma_{{\rm A},ij},
\end{align}
where 
\begin{align}
\sigma_{{\rm S},ij}=-\Pi\delta_{ij}+(\eta_{\rm d}-\eta_{\rm s})\delta_{i j}\partial_kv_k
+\eta_{\rm s}(\partial_j v_i+\partial_i v_j),~~~~
\sigma_{{\rm A},ij}=\frac{1}{2}\eta_{\rm o}
\left( \partial_j v_i^\ast+\partial_i v_j^\ast+\partial_j^\ast v_i+\partial_i^\ast v_j \right),
\end{align}
with $\sigma_{ij}=\sigma_{{\rm S},ij}+\sigma_{{\rm A},ij}$.
Integrating the above equation over the fluid area $A$, we obtain the integral identity as
\begin{align}
&\int_C {\rm d}s\, 
n_i (\sigma^\prime_{{\rm S},ij}+\sigma^\prime_{{\rm A},ij}) v_j
-\int_C {\rm d}s\, n_i(\sigma_{{\rm S},ij}-\sigma_{{\rm A},ij})v_j^\prime \nonumber\\
&=  -\int_A {\rm d}A\,  v_j^\prime b_j + \int_A {\rm d}A\, v_jb_j^\prime 
+ \int_A {\rm d}A\, \Pi^\prime\partial_jv_j  -\int_A {\rm d}A\, \Pi\partial_jv_j^\prime -2 \int_A {\rm d}A\, v_j^\prime \partial_i \sigma_{{\rm A},ij},
\label{eq:NHT}
\end{align}
where $C$ denotes the curve bounding the area $A$ and the unit vector $\mathbf{n}$ is directed into that area.
We note that Eq.~(\ref{eq:NHT}) is not invariant under the exchange of the unprimed and primed variables when $\eta_{\rm o}\neq0$.
When $\eta_{\rm o}=0$, on the other hand, Eq.~(\ref{eq:NHT}) reduces to the Lorentz reciprocal theorem for a 2D compressible fluid~\cite{cunha2003}.

\section{Derivation of Eq.~(\ref{eq:noslip})}
\label{app:BIE}

Here we derive the boundary integral equation in Eq.~(\ref{eq:noslip}).
When $\Pi=\Pi^\prime=0$, we consider the following two types of unprimed and primed flows
\begin{align}
\partial_j \sigma_{ij}  = 0,~~~~ \partial_j \sigma_{ij}^\prime + F_i^\prime\delta(\mathbf{R}-\mathbf{R}^\prime) = 0.
\label{eq:2flows}
\end{align}
Suppose that a circular rigid disk is moving in the fluid area $A$, Eq.~(\ref{eq:NHT}) with the use of Eq.~(\ref{eq:2flows}) becomes~\cite{pozrikidis1992,masoud2019}
\begin{align}
\int_{C_{\rm d}} {\rm d}s(R)\, n_i(\sigma_{{\rm S}, ij}-\sigma_{{\rm A}, ij})v_j^\prime 
= \int_{C_{\rm d}} {\rm d}s(R)\, 
n_i (\sigma_{{\rm S}, ij}^\prime+\sigma_{{\rm A},ij}^\prime) v_j
- F_j^\prime\int_A {\rm d}A(R)\,  v_j \delta(\mathbf{R}-\mathbf{R}^\prime),
\end{align}
where $C_{\rm d}$ is the circular curve bounding the moving disk, as schematically depicted in Fig.~\ref{fig:app}.
The notations, ${\rm d}s(R)$ and ${\rm d}A(R)$, indicate that $\mathbf{R}$ is the integration variable.
Assuming a non-slip boundary condition at the disk perimeter and using the form of the point-force solution, $v_i^\prime(\mathbf{R}) = G_{ij}(\mathbf{R}-\mathbf{R}^\prime)F_j^\prime$, we obtain~\cite{masoud2019}
\begin{align}
\int_A {\rm d}A(R)\,  v_\ell(\mathbf{R}) \delta(\mathbf{R}-\mathbf{R}^\prime)
=
-\int_{C_{\rm u}} {\rm d}s(R)\, f_j(\mathbf{R})G_{j\ell}(\mathbf{R}-\mathbf{R}^\prime),
\label{eq:IBE}
\end{align}
where $C_{\rm u}$ denotes the domain of the unknown force distribution, $\mathbf{f}$, and we have assumed that the force distribution can be defined as $f_j=n_i(\sigma_{{\rm S}, ij}-\sigma_{{\rm A}, ij})$.
Interchanging $\mathbf{R}$ and $\mathbf{R}^\prime$, we finally obtain Eq.~(\ref{eq:noslip}).

\begin{figure}[tbh]
\begin{center}
\includegraphics[scale=0.5]{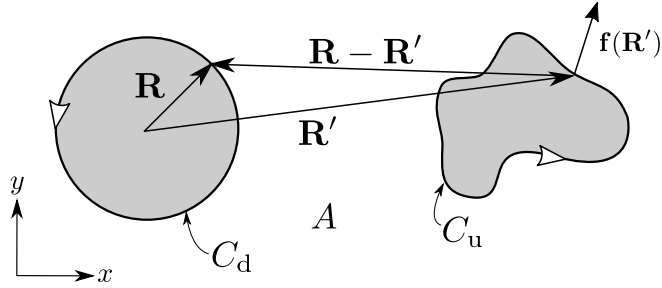}
\end{center}
\caption{
Sketch of the circular curve, $C_{\rm d}$, and the unspecified curve, $C_{\rm u}$, with the accompanying unknown force distribution, $\mathbf{f}$, while $A$ is the fluid area bounded by both $C_{\rm d}$ and $C_{\rm u}$ (white area).
The curves, $C_{\rm d}$ and $C_{\rm u}$, are parameterized by the vectors $\mathbf{R}$ and $\mathbf{R}^\prime$, respectively and the two arrows represent the direction of the line integral.
In the sketch, we have $\vert \mathbf{R}^\prime \vert > \vert \mathbf{R} \vert$ only for presentation purposes.
In actual calculations where the condition $\vert \mathbf{R}^\prime \vert \ll \vert \mathbf{R} \vert$ is used, the two curves overlap with each other.
}
\label{fig:app}
\end{figure}

\section{Derivation of Eqs.~(\ref{eq:translation}) and (\ref{eq:rotation})}
\label{app:resistance}

We derive the translational friction tensor $\mathbf{\Gamma}$ and the
rotational friction coefficient $\Lambda$ in Eqs.~(\ref{eq:translation}) and (\ref{eq:rotation}),
respectively.
If we assume $\vert \mathbf{R}^\prime \vert \ll \vert \mathbf{R} \vert$~\cite{oppenheimer2009}, the right-hand-side
of Eq.~(\ref{eq:noslip}) can be expanded up to first order in $\mathbf{R}^\prime$ as
\begin{align}
U_{i}+\epsilon_{ijk}\Omega_jR_k
\approx -
\int_{C_{\rm u}} {\rm d}s(R^\prime)\,
f_{j}\left(\mathbf{R}^{\prime}\right)
\left[
G_{ji}(R) - R_k^\prime \frac{\partial G_{ji}(R)}{\partial R_k}
\right].
\label{eq:noslipa}
\end{align}
Integrating Eq.~(\ref{eq:noslipa}) over the circular disk perimeter, ${C_{\rm d}}$,
parametrized by $\mathbf{R}$, we obtain the relation between the velocity and the force as
\begin{align}
U_i = -\frac{1}{2\pi R}
\int_{C_{\rm u}} {\rm d}s(R^\prime) \, f_j(\mathbf{R}^\prime)
\int_{C_{\rm d}} {\rm d}s(R) \, G_{ji}(R)
= -[(C_1+C_2/2)\delta_{ij} - C_3\epsilon_{ij}]F^{\rm d}_j,
\end{align}
where $F^{\rm d}_i=\int_{C_{\rm u}} {\rm d}s(R^\prime)\, f_i(\mathbf{R}^\prime)$ is the force acting on the disk.
In the above, the integrals of the odd terms in $\mathbf{R}$ vanish because of the symmetry
of the disk.
Hence, we obtain Eq.~(\ref{eq:translation}).

Next, we multiply both sides of Eq.~(\ref{eq:noslipa}) by $\mathbf{R}$ and integrate
over $C_{\rm d}$
\begin{align}
\pi R^3\epsilon_{\ell ij}\Omega_j
=
\int_{C_{\rm u}} {\rm d}s(R^\prime)\, R_k^\prime f_{j}\left(\mathbf{R}^{\prime}\right)
\int_{C_{\rm d}} {\rm d}s(R)\,
 R_\ell \frac{\partial G_{ji}(R)}{\partial R_k}.
\label{eq:torque}
\end{align}
We further multiply both sides of the above equation by $\epsilon_{\ell in}$ and obtain
\begin{align}
\Omega_n = \frac{1}{2R}\left[\left( \frac{\partial C_1}{\partial R} -\frac{C_2}{R}\right)\epsilon_{nkj} - \frac{\partial C_3}{\partial R}\epsilon_{nki}\epsilon_{ij}\right] 
\int_{C_{\rm u}} {\rm d}s(R^\prime)\, R_k^\prime f_{j}\left(\mathbf{R}^{\prime}\right).
\end{align}
On the right-hand-side, the term with $C_3$ represents the radial pressure acting on the disk~\cite{avron1998}.
As this term does not contribute to the torque, it vanishes and the relation between the angular
velocity and the torque becomes
\begin{align}
\mathbf{\Omega} &= \frac{1}{2R}\left( \frac{\partial C_1}{\partial R} -\frac{C_2}{R}\right)\mathbf{T}^{\rm d},
\end{align}
where $T^{\rm d}_i=\epsilon_{ijk}\int_{C_{\rm u}} {\rm d}s(R^\prime)\, R_j^\prime f_k\left(\mathbf{R}^{\prime}\right)$ is the torque on the disk.
Hence, we obtain Eq.~(\ref{eq:rotation}).

\end{widetext}


%

\end{document}